# Simple experiment to test a hypothetical transient deviation from Quantum Mechanics

Alejandro A. Hnilo and Mónica B. Agüero

*CEILAP, Centro de Investigaciones en Láseres y Aplicaciones, CITEDEF, J.B. de La Salle 4397, (1603) Villa Martelli, Argentina.*
email: alex.hnilo@gmail.com

April 15th, 2019.

Quantum Mechanics (QM) predicts the correlation between measurements performed in remote regions of a spatially spread entangled state to be higher than allowed by the intuitive concepts of Locality and Realism (LR). This high correlation forbids the introduction of nonlinear operators of evolution in QM (which would be desirable for several reasons), for it would lead to faster-than-light signaling. As a way out of this situation, it has been hypothesized that the high quantum correlation can be observed only after a time longer than $L/c$ has elapsed (where $L$ is the spatial spread of the entangled state and $c$ is the speed of light). In shorter times, a level of correlation compatible with LR would be observed instead. A simple model following this hypothesis is described. It has not been disproved by any of the performed experiments to date. A test achievable with accessible means is proposed. The data recorded in a similar but incomplete experiment (which was done in 2012 with a different purpose, and repeated in 2018 producing essentially the same results) are analyzed, and are found consistent with the described model. Yet, we stress that a specific experiment is absolutely needed.

**1. Introduction.**

The Copenhagen interpretation of Quantum Mechanics (QM) has faced debate since its early years. At least three points of conflict have been identified:
*i)* The failure of the principle of correspondence in chaotic systems ("quantum chaos").
*ii)* The evolution from a superposition state to the observed state of a single system is not described by the theory ("measurement" or "projection" problem).
*iii)* The correlation between measurements performed in remote regions of a spatially extended entangled state is higher than allowed by Local Realism.

Local Realism (LR) is a shorthand for the intuitive notions of the separability of physical phenomena, and that the properties of the world are independent of being observed or not [1,2]. LR is assumed not only in everyday life, but in the practice of all scientific fields (excepting QM).

Conflictive points *(i)* and *(ii)* may be solved using nonlinear operators of evolution [3]. But, it has been shown [4,5] that such operators would allow, if combined with the high correlation of point *(iii)*, faster-than-light signaling. This would be in contradiction with theory of Relativity, and make quantum field theory untenable. The conclusion is that a nonlinear extension of QM is impossible, even if it is infinitesimal. This conclusion has been interpreted in two opposite ways. For some, it is the demonstration that QM is part of the "ultimate theory", for the introduction of correction terms is impossible. For others, instead, it means that the predictions of QM are structurally unstable, and hence that there is something important missing in its formulation.

Hope to reconcile QM with LR come up by noting that the performed tests leave space to the following hypothesis: QM predictions are valid to statistical averages measured over "long" times; measurements performed in "short" times would hold to LR. Timescale here would be given by the time light needs to cover the spatial spread $L$ of the entangled state. In other words: QM (as we know it) would describe t→∞ stationary states, while transient deviations from QM predictions would be observed in times shorter than $L/c$. If this hypothesis was demonstrated correct, it would solve not only point *(iii)*, but it would also open the door to the use of nonlinear operators to solve points *(i)* and *(ii)*. Faster-than-light signaling would never be possible, for the high correlation characteristic of entanglement would arise only after a time longer than $L/c$ has elapsed. Note that usual QM would still make correct predictions in the overwhelming number of cases. The only case where we foresee some limitation is Quantum Key Distribution. In principle, transient deviations would impose the rate of detected entangled pairs to be smaller than $c/L$, a limit which is anyway many orders of magnitude above what can be reached nowadays.

Following that hypothesis, we introduce in this paper a hidden variables model named "AB". It can be tested in a simple experiment. In the next Section 2, AB is described. As almost all such models, it is an artificial construction. Yet, it is expected to reveal, at least, the essential features of the hypothesized transient deviations. In the Section 3, the appropriate testing experiment is explained. The case of an asymmetrical setup is briefly studied in Section 4. In the Section 5, we analyze data obtained in a similar but incomplete experiment performed in 2012 and repeated (with essentially the same results) in 2018. These data are consistent with AB, but we do not claim them to provide any sort of confirmation. We think they just encourage performing the appropriate experiment.

**2. The AB model.**
*2.1 A simple hidden variables model.*

Because of its relevance and simplicity, the case of two photons entangled in polarization is considered in what follows, see Figure 1. Let assume that a source S emits pairs of photons in the fully symmetrical Bell state: $|\varphi^+\rangle = (1/\sqrt{2})\{|x_A,x_B\rangle+|y_A,y_B\rangle\}$ towards distant stations A and B. Correlation is evaluated, for example, with the Clauser-Horne-Shimony and Holt parameter $S_{CHSH}$ [1]. It involves measuring the probability of double coincidences at the two transmitted ports of the analyzers $P^{++}(a,b)$, at the two reflected ports $P^{--}(a,b)$, and the mixed cases $P^{+-}(a,b)$ and $P^{-+}(a,b)$, with angle settings $a = \{0,\pi/4\}$ and $b = \{\pi/8, 3\pi/8\}$. The QM ideal prediction is $S_{CHSH} = 2\sqrt{2}$, while LR imposes $S_{CHSH} \leq 2$.

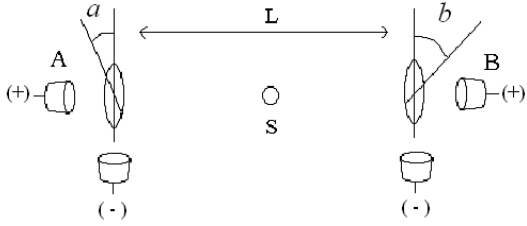

Figure 1: The source S emits entangled states $|\varphi^+\rangle$ towards stations placed at a distance $L$. Analyzers are set at angles $a$ and $b$, and single photons are detected transmitted (+) or reflected (-) at each station.

An important parameter is the *efficiency* $\eta \equiv N_c/N_s$, where $N_c$ is the number of coincidences and $N_s$ the number of single detections (both recorded in a certain time). Note that $\eta \leq 1$.

Let assume now that each emitted pair carries an angular "hidden variable" $\alpha$ that takes values in $[0,2\pi]$ and has two equally probable sub-states $\alpha^+$ and $\alpha^-$. The probabilities of detection at station A (setting $a$) is:

if $\alpha = a$ and $\alpha=\alpha^+$, $P_A^+ = 1$ and $P_A^- = 0$;
if $\alpha = a$ and $\alpha=\alpha^-$, $P_A^+ = 0$ and $P_A^- = 1$;    (1a)
if $\alpha \neq a$, $P_A^+$ and $P_A^- = 0$.

and at station B:

if $\alpha=\alpha^+$, $P_B^+ = cos^2(b-\alpha)$ and $P_B^- = sin^2(b-\alpha)$
if $\alpha=\alpha^-$, $P_B^+ = sin^2(b-\alpha)$ and $P_B^- = cos^2(b-\alpha)$    (1b)
(regardless whether $\alpha = a$ or not, for this is unknown at station B).

where $P_i^+$ ($P_i^-$) is the probability to detect a photon at detector "+" ("-") of station $i$. The probability of coincidence in the "+" ports is then:

$$P^{++} = p(\alpha=a) \times \tfrac{1}{2} \times cos^2(b-a) \quad (2)$$

where $p(\alpha=a)$ is the probability that $\alpha=a$ (the ½ comes from the two equally probable sub-states of $\alpha$). This result is proportional to the QM prediction. The same holds for the coincidence probabilities $P^{+-}$, $P^{-+}$ and $P^{--}$, so that $S_{CHSH} = 2\sqrt{2}$. However, note that there is an asymmetry in the number of single counts in each station: "all" pairs are detected at B, while only "some" of them are detected at A. To be precise: if $R^\alpha$ is the total rate of photons (regardless if + or -) detected at B, the rate detected at A is only $p(\alpha=a) \times R^\alpha$. In order to balance the rates detected at each station, let assume that half of the emitted pairs carry a hidden variable $\beta$ that mirrors eqs.1:

if $\beta = b$ and $\beta=\beta^+$, $P_B^+=1$ and $P_B^- = 0$;
if $\beta = b$ and $\beta=\beta^-$, $P_B^+= 0$ and $P_B^- = 1$;    (3)
if $\beta \neq b$, $P_B^+$, $P_B^- = 0$.

and:

if $\beta=\beta^+$, $P_A^+ = cos^2(a-\beta)$ and $P_A^- = sin^2(a-\beta)$    (3b)
if $\beta=\beta^-$, $P_A^+ = sin^2(a-\beta)$ and $P_A^- = cos^2(a-\beta)$

Choosing $p(\alpha=a) = p(\beta=b) \equiv p$, the same rate is detected at each station. The probability of detecting a single photon at A is: *probability of being an $\alpha$-pair $\times$ $p(\alpha=a)$ + probability of being a $\beta$-pair $\times$ 1* (for, all $\beta$-pairs are detected at A) = ½ $p$ + ½ . The probability of detecting a coincidence is ½$\times p(\alpha=a)$ + ½$\times p(\beta=b) = p$. Therefore, the efficiency in this symmetrical case is:

$$\eta = 2p / (1 + p) \quad (4)$$

In the case of experiments with fast variation of analyzers' settings [6,7] there are two possible settings in each station, so that $p = ½ \Rightarrow \eta = ⅔$. This is the well-known result that $\eta > ⅔$ is necessary to disprove LR, because of the efficiency loophole [1]. For this reason, great efforts have been made to improve efficiency [8]. In the general case, it is convenient to relax the strict equalities in eqs.1 and 3 to:

$P_A^+$ or $P_A^- = 1$ if $\alpha \in [a - ½\Delta, a + ½\Delta]$
$P_A^+$ and $P_A^- = 0$ otherwise.    (5)
$P_B^+$ and $P_B^-$, as in eqs.1b
(and symmetrically for $\beta$).

Now $p = \Delta/2\pi$. After integrating $\alpha$ and $\beta$ in $[0,2\pi]$, the total probability of "++" coincidence is:

$$P^{++} = (1/4\pi)\{sin(\Delta)\times cos^2(a-b) + ½ [\Delta - sin(\Delta)]\} \quad (6)$$

If $\Delta<<1$, $P^{++} \approx (\Delta/4\pi)\times cos^2(a-b) = p\times½\times cos^2(a-b)$ as before. If $\Delta=2\pi$ then $p=1$ and $\eta=1$, but $P^{++}= ¼$ (no correlation). The other three coincidence probabilities are analogous to eq.6. The result is that, for the usual setting angles:

$$S_{CHSH} = 2\sqrt{2}\, sin(\Delta)/\Delta \quad (7)$$

Be aware that eqs.6 and 7 hold if there is no correlation between the values of $\alpha$ and $a$ ($\beta$ and $b$). An upper limit to the possible value of $\Delta$ in Nature can be estimated from measured values of $S_{CHSH}$. Using $S_{CHSH}$ = 2,73 [7], $\Delta < 0,45$.

A pertinent question is whether the AB model is disproved by the recent loophole-free experiments, or not. Direct comparison is impossible, for these experiments use, in some cases, Eberhardt's states and Clauser-Horne inequality instead of $|\varphi^+\rangle$ and $S_{CHSH}$ [9-

10] and, in other cases, entanglement swapping between photons and atoms [11-12]. A sort of critical review of these experiments can be found in [13]. The question deserves a detailed discussion, but note that what is important about AB is not the precision it fits the QM predictions, but what it predicts *different* from QM. For the purposes here, it suffices to say that no test based on measuring *time averaged* magnitudes is able to rule out the AB model. As shown in the next section, all (averaged) QM predictions are reproduced by AB as t→∞.

*2.2 Assume a delayed back reaction.*

Up to this point, nothing really new has been introduced. Let assume now that the detection of a photon at A produces a "reaction" on the field. This reaction propagates backwards to the source S, carrying information on properties the photon had when it was detected (in the case of interest here, whether it was detected polarized with an angle parallel, or perpendicular, to *a*). The reaction arrives to S a time $\tau \geq L/c$ after the emission ($\tau/2$ to go and $\tau/2$ to return), and influences the values of α emitted thereafter, in such a way that they tend to fit α=*a* (a sort of stimulated emission at the hidden variables level). The reaction is proportional to the rate of detected photons $R_{det}$ and to the mismatch between α and *a* when the photon was detected (that is, when the reaction was born). Therefore, the evolution of the values of α emitted at time t is given by:

$$d\alpha(t)/dt \propto - R_{det}(t-\tau/2).[\alpha(t-\tau) - a(t-\tau/2)] \qquad (8)$$

but $R_{det}(t-\tau/2) \propto R_{emitted}(t-\tau)$, so that:

$$d\alpha(t)/dt = - \Gamma(t-\tau).[\alpha(t-\tau) - a(t-\tau/2)] \qquad (9)$$

where the value of Γ is unknown, but is supposedly proportional to the rate of emitted photons (which is, in turn, proportional to the pump intensity). Before the delayed reaction arrives, S emits random values of α. If S emits entangled states, both beams carry the same value of α. If S is prepared to emit classical states instead, their values of α are not correlated. A similar equation holds for β(t).

The reaction that is hypothesized here is inspired by the absorber theory of radiation proposed by Wheeler and Feynman [14]. In this theory, the reaction propagates backwards not only in space but also in time. D.Pegg shows that this theory explains the experiment in Fig.1 in an elegant way [15]. This is known now as the *transactional interpretation* of QM. The theory is impeccable, but the idea of signals propagating backwards in time is uncomfortable. The "normal time" reaction hypothesized here has a poorer theoretical basis, but it is free of that uncomfortable feature and, most important, it leads to a testable difference with QM.

Eq.9 is a delay differential equation. It evolves in a phase space of infinite dimensions. Its solutions are difficult to find in the general case. In the particular case that Γ(t) and *a*(t) are constant, the solutions have general form α(t)= A×exp(zt/τ), z∈C. Critical damping of α(t) towards *a* is reached for Γτ =1/e. For 1/e< Γτ < π/2, α(t) converges towards *a* with damped oscillations of period ≈4τ. $S_{CHSH}$ is always given by eq.(7), while η(t) depends on Δ and the distance between α(t) and *a*. As t→∞, then α→*a*, β→*b*, p→1, η→1, $S_{CHSH}$ → 2√2, *all* observables coincide with QM predictions regardless the value of Δ. On the other hand, the amplitude of oscillations of α(t) diverges if Γτ ≥ π/2. In this case α(t) runs away losing any correlation with *a*. No steady state is reached. Eq.(7) for $S_{CHSH}$ still holds, but η ≤ ⅔. The same applies to the experiments performed using random and fast variation of analyzers' settings.

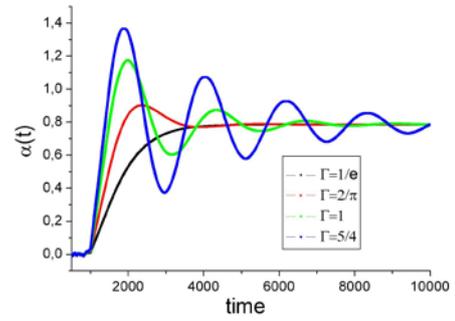

Figure 2: α(t) for several values of Γ scaled with τ (note that plot starts at t=τ) according to eq.9; Γ(t<0)=0, Γ(t≥0)=Γ, α(t<τ)= random (not plotted), α(t=τ)=0, *a*=π/4, timescale: τ=500. For critical damping (Γ=1/e), the value 0,9×*a* is reached in ≈3,5 τ.

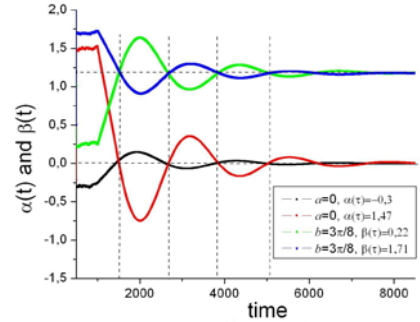

Figure 3: An example of spontaneous synchronization. The evolution of α(t) and β(t) are plotted for different initial conditions (including 500 different random values for t<τ, not plotted); and target values (*a*=0, *b*=3π/8), Γ=1. In spite of all these differences, target values are crossed at nearly the same times. Synchronization improves as the partition of τ increases (here, τ=500).

To explore the behavior of α(t) further, we run numerical simulations of eq.9. Timescale is given by τ, divided into 500 smaller intervals (this number is found by increasing the number of intervals until the results stabilize). In Figure 2, Γ is "turned on" at t=0 and remains constant thereafter; we choose *a* = π/4 and α(t=τ) = 0. Values of α(t) for t<τ are random in [0,2π], they are not plotted to not to burden the figure. At t=τ the reaction starts to influence the emitted values of α.

Yet, as the values $\alpha(t-\tau)$ in the rhs of eq.9 are randomly distributed, $\alpha(t)$ does not evolve far from $\alpha(\tau)$ during this stage. At $t=2\tau$ the values of $\alpha(t-\tau)$ are not random any longer, and $\alpha(t)$ starts to evolve towards the "target" value $a$. It does so in a monotonous or oscillatory way, depending on the value of $\Gamma$.

An interesting numerical result is that oscillations cross their target values (0 and $\pi/4$) at nearly the same time (if $\Gamma$ is constant), even if initial values $\alpha(t\leq\tau)$ are all different. The only requisite for this to occur is that the random initial values are distributed symmetrically to the target. As the former are uniformly distributed in $[0,2\pi]$, the requisite is fulfilled for any target. This effect of *spontaneous synchronization* (Figure 3) has important consequences, as it is shown later.

In the next Section, a simple experiment to test model AB is described.

## 3. A proposal to test AB.
*3.1 Description of the setup.*

According to QM, there is no reason for a time variation of the efficiency. An elementary approach to test QM vs AB is hence to measure $\eta(t)$ in a time shorter than $\tau$. Yet, this experiment is difficult to perform. In order to check entanglement and record $\eta(t)$, a rate of coincidences $>10^3$ $\tau^{-1}$ is required. The highest reported rate is $\approx 3\times 10^5$ s$^{-1}$ in a laboratory environment [16], what means $\approx 10^{-4}$ $\tau^{-1}$. Numbers are better for larger $L$: 50 s$^{-1}$ at 13 Km [17] ($\approx 2\times 10^{-3}$ $\tau^{-1}$) and 8 s$^{-1}$ at 144 Km [18] ($\approx 4\times 10^{-3}$ $\tau^{-1}$), but still short of required by almost six orders of magnitude. Nevertheless, it is reasonable to expect $\alpha(t)$ to decay to a random distribution after the driving reaction is turned off. This decay should take a finite time $\tau_d$. A "stroboscopic" measurement of $\eta(t)$ is then possible, by using a pulsed source of pairs with time between pulses longer than $\tau_d$. Although $\tau_d$ is unknown, pulse repetition rate $R_p$ can be adjusted until some effect is perceived. If $\tau_d$ turns out to be long the detections' rate may be forced to be low, but this is not an unsolvable problem, for satisfactory statistics can be attained by indefinitely increasing the number of recorded pulses.

Let detail the proposed experiment:
*i) The source of pairs*: A set of nonlinear crystals is pumped with square laser pulses of risetime and fall time $\tau_{rf}$ and full duration $\tau_{pulse}$, equally separated by a time $R_p^{-1}$. All these parameters, and also the pump intensity, are adjustable. A laser diode at 405 nm pumping a pair of two crossed type-I phase matched nonlinear crystals [19] seems to be the simplest choice.
*ii) The stations A and B*: are provided with devices to record time of detection of each photon, and also time of emission of each pumping pulse, with resolution $\tau_{res}$. The assumed spontaneous decay of $\alpha(t)$ and $\beta(t)$ to the random state implies that fast and unpredictable variation of analyzers' settings $\{a,b\}$ is unnecessary. This is not only an important practical simplification, but it also solves the problem of ensuring the settings' unpredictability, which is an involved issue [20-22].

Distance $L$ is adjustable in order to scan timescale $\tau=L/c$. Pairs' propagation through optical fibers (as in [7]) is advisable to avoid spurious variations f $\eta$ with $L$ due to possible changes in alignment.
*iii) Time hierarchy*: the following relationships should hold:
$\tau_{res} < \tau$, to resolve details of $\eta(t)$.
$\tau_{rf} \ll \tau$, to pump pulse to be square shaped.
$\tau \ll \tau_{pulse}$, to give enough time to the evolution of $\alpha(t)$.
$\tau_{pulse} \ll R_p^{-1}$, to pulses to be well separated.
$\tau_d \ll R_p^{-1}$, to allow $\alpha(t)$ to decay to random state.

The first critical time in this hierarchy is $\tau_{res}$. Time-stamping devices are claimed to provide picosecond resolution, but jitter of avalanche photodiodes (usual for single photon detection) limit $\tau_{res}$ to $\approx 2$ ns. Standard laser diodes and driving devices also reach $\tau_{rf} \approx 2$ ns. These numbers imply values for $\tau$ in the 20-200 ns range ($L$ from 6 to 60 m), and hence $\tau_{pulse}$ from 0.2 to 2 $\mu$s, and $R_p$ from 500 to 50 KHz. All these figures are easily attainable. Recall that $R_p$ must be explored below these numbers, because $\tau_d$ is unknown.

| Time parameter (name) | Description |
|---|---|
| $\tau$ | Time between stations, $L/c$ |
| $\tau_{res}$ | Resolution of time stamping |
| $\tau_{rf}$ | Pulse rise and fall time |
| $\tau_{pulse}$ | Pulse duration |
| $\tau_d$ | Decay time to initial state |
| $R_p^{-1}$ | Separation between pulses |

Table: Summary of time parameters involved in the proposed experiment. All of them are adjustable by the observer excepting $\tau_d$, whose value is unknown.

Pump intensity must be kept low, so that probability of detection of one photon per pulse remains small. This is to limit the number of spurious coincidences, which is a serious issue in the pulsed regime [23]. As a consequence, one should expect a single photon rate in the 5-50 KHz range.

*3.2 What to do.*

A sketch of the protocol:
#1) Place stations at a distance $L$ of each other and record coincidences at different values of $\{a,b\}$ to check that entangled states are actually being observed. From time stamped files of photons' detections and pulses' emissions, calculate $\eta(t)$.
#2) Change values of pump intensity, pulse duration, and repetition rate.
#3) Move the stations at a different distance $L'$, and repeat steps #1 and #2.
#4) If a time evolution of $\eta$ is in fact observed, check if it depends on the straight line distance between the stations, or on the distance measured through the optical fibers. This would reveal whether the "reaction" propagates backwards following the photons' paths.

We run numerical simulations of AB, assuming that $\Gamma$ is turned on at $t=0$ and remains constant thereafter.

The sharp "slit" assumed in eq.5 is simple but rather unphysical; in these simulations we use instead the smooth condition:

$$p(\alpha=a) = exp\ -[\alpha(t)-a]^2/\Delta^2 \qquad (10)$$

(and the same for B, β, b). From eq.4, then:

$$\eta_A(t) = \{exp\ -(\alpha-a)^2/\Delta^2 + exp\ -(\beta-b)^2/\Delta^2\} / \{1 + exp\ -(\alpha-a)^2/\Delta^2\} \qquad (11)$$

(and a symmetrical expression for $\eta_B$), where α and β are functions of time ruled by eq.9.

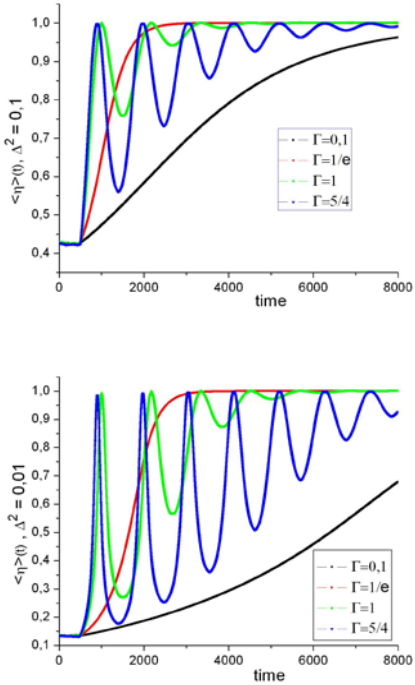

Figure 4: observable <η>(t) averaged over $10^3$ "pulses" with different random initial conditions and values of Γ (scaled with τ), $\Delta^2 = 0,1$ (up), $\Delta^2 = 0,01$ (down), timescale: τ=500.

To take into account that α(t) and β(t) are in a random state before the arrival of the reaction, each numerical iteration (each iteration corresponds to a pump pulse) starts from a different set of random initial conditions. There is no memory from one "pulse" to the next. After many such pulses, it is obtained a curve of the efficiency averaged over a large number of different sets of initial conditions, <η>(t). The right number of pulses is found by increasing it until the curve of <η>(t) stabilizes. In all tested cases, $10^3$ pulses suffice. In spite of the explicit functional dependence in eq.11, numerically obtained values of <η>(t) are found to be independent of {a,b}. This is the consequence of the spontaneous synchronization effect mentioned before. Some results of numerical simulations are displayed in Figures 4. Note that, in order to draw these curves, an arbitrarily high rate of pairs is implicitly supposed during each "pulse", although the value of Γ remains (of course) finite.

For Δ fixed, the curves start with the same value of <η>(t=τ) regardless the value of Γ. This is obvious because the reaction, no matter how weak or strong, has had no time to propagate back from the stations to the source. Naturally, <η>(t=τ) is smaller for smaller Δ (which means a narrower "slit"). As in Fig.2, curves remain near <η>(t=τ) for a time τ (i.e., until the earliest reaction arrives to the source). Then α and β start to evolve towards their target values *a* and *b*, and the curves split depending on the value of Γ. The evolution is faster, and the saturation value (=1 in these ideal conditions) is reached sooner, for larger Γ.

If the value of Γ is such as to produce oscillations, a peak of efficiency is reached each time α(t) and β(t) are close to a,b. These peaks are visible in Figs.4 in spite of the different initial conditions and target values, thanks to the spontaneous synchronization effect. For Γ small instead, a nearly linear increase of <η>(t) is visible. If Δ is relatively large, the efficiency saturates sooner and the concavity of the curve is downwards. If Δ is small, the concavity is upwards. For Γ large, oscillations are sharper and show higher contrast as Δ decreases, but the period remains the same.

*3.3 What to observe, and approximate expressions.*

If AB is correct, the following phenomena should be then observed:

*I) At fixed L*: <η>(t) increases with time until it saturates. Time needed to reach the saturation value decreases if pump intensity is increased. At high levels of pump intensity, oscillations or "peaks" may appear. If $R_p$ is increased above a certain threshold (the unknown value $\tau_d^{-1}$), <η>(t=0) would also increase, for α(t) and β(t) would have had no time to decay.

*II) At variable L*: the time <η>(t) takes to reach saturation and the period of the oscillations increase with *L*.

These features mean a test of QM vs LR different from all the ones attempted until now. It is independent of the violation of a statistical correlation limit. It is also more robust, for a definite result is obtained even if the experimental setup is far from ideal regarding efficiency of optics, detectors and alignment. The key is the scaling of the dynamics of <η>(t) with *L*.

It is useful having analytical expressions, even if they are only approximate. For Γτ ≤ 1/e (no oscillations):

$$\alpha(t) \approx [a - \alpha(\tau)] \times \{1 - exp[-\Gamma \times (t-\tau)]\} + \alpha(\tau) \qquad (12)$$

Due to spontaneous synchronization, exponentials in eq.11 are maximal at the same time: $exp\ -(\alpha-a)^2/\Delta^2 \approx exp\ -(\beta-b)^2/\Delta^2$ and $\eta_A(t) = \eta_B(t) \equiv \eta(t)$. Then eq.11 can be approximated as:

$$\eta(t) \approx exp\ -[\alpha(t) - a]^2 / \delta^2 \qquad (13)$$

where $\delta = [\sqrt{(ln3)/(2ln2)}]\ \Delta \approx 0.7561\ \Delta$. Using eq.12:

$$\eta(t) \approx exp\ -(\{[\alpha(\tau)-a]^2 exp[-2\Gamma \times (t-\tau)]\}/\ \delta^2) \qquad (14)$$

averaging over the (random) initial conditions α(τ) and assuming δ<<1:

$$<\eta>(t) \approx \sqrt{\pi} \times \delta \times exp[\Gamma \times (t-\tau)] \quad (15)$$

and hence the efficiency and its slope at t=τ are:

$$<\eta> \approx 1{,}34\,\Delta \quad (16)$$

$$d<\eta>/dt \approx 1{,}34\,\Gamma \times \Delta \quad (17)$$

These expressions fit the results of numerical simulations satisfactorily. F.ex., for Fig.4a $<\eta>(t=\tau) = 0{,}43$ while according to eq.16 $<\eta>(t=\tau)= 0{,}42$. Eqs.16-17 may be a useful tool to quickly get the order of magnitude of the main parameters of the model from measured data, as it is illustrated in the Section 5.

## 4. The case of an asymmetrical setup.
*4.1 Answering a question.*

Dr. Siddarth Koduru Joshi (from IQOQI) put forward the interesting and fruitful question of what would be observed (according to the AB model) if the stations in Fig.1 are placed at different distances from the source. Now two timescales are relevant: $\tau_A = L_A/c$ and $\tau_B = L_B/c$, where $L_A$ ($L_B$) is the distance from the source to A (B) station. There are also two possible interaction strengths: $\Gamma_A$ and $\Gamma_B$. In the previous, symmetrical case, Γ is scaled with τ, and hence one gets a single-parameter set of solutions. In the non-symmetrical case, the other interaction strength and time delay have independent values, and hence one gets a three-parameters set of solutions. An exhaustive exploration of the 3-D parameter space is impossible to display here. Just a couple of relevant cases are presented.

*4.2 $L_A$, $L_B$ are slightly different.*

In Figure 5, the time evolution of coincidences and observable efficiencies at stations A and B are displayed for the case $L_A/L_B = 5/6$, Γ=1.25 for both arms (be aware of the different scaling of Γ for each branch), $a = 0$, $b = \pi/8$, $\Delta^2 = 0.1$.

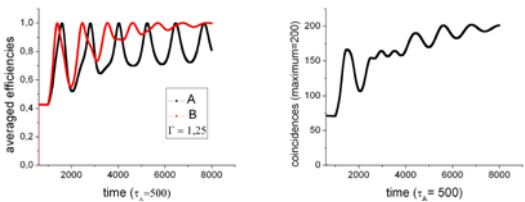

Figure 5: $L_A/L_B = 5/6$. Left: observable $<\eta_A>$(t) and $<\eta_B>$(t) averaged over 100 "pulses" with different initial conditions, $a=0$, $b = \pi/8$, $\Delta^2 = 0.1$. Right: number of coincidences.

Note the deformation of the efficiencies' oscillations in the early periods (i.e., they are not harmonic). A curious result is that the efficiency of the station which is farther from the source reaches the maximum value (=1) first, and that $\eta_A$(t) oscillates with the period that corresponds to the B arm (and vice versa). In this case the spontaneous synchronization effect holds, so that the efficiencies are independent of the values of the angle settings *a,b* (note that both curves start at the same efficiency value). The evolution of the number of coincidences (which, of course, is the same for both stations) is beating-like, as could be expected (Fig.5 on the right).

*4.3 $L_A$, $L_B$ are very different.*

In Figure 6, the time evolution of coincidences and observable efficiencies at stations A and B are displayed for the case $L_A/L_B = 1/7.5$. All the other parameters are the same than in section 4.2

The efficiency in the station nearest to the source (A) follows a fast oscillation (i.e., at time scale $\tau_A$) and then slower oscillations (i.e., at time scale $\tau_B$). The efficiency in the farthest station (B), instead, rapidly converges to the maximum value. Note that in this case the curves start at different points, i.e., $\eta_A(t=\tau_A) \neq \eta_B(t=\tau_B)$. This indicates that the spontaneous synchronization effect *no longer holds*. In consequence the efficiencies can be functions of the angle settings *a,b*, what would be easy to observe. Finally, the time evolution of the number of coincidences shows the same "fast" and "slow" time scales, and is practically identical to $\eta_A$(t).

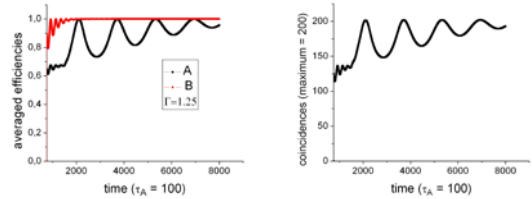

Figure 6: $L_A/L_B = 1/7.5$. Left: observable $<\eta_A>$(t) and $<\eta_B>$(t) averaged over 100 "pulses" with different initial conditions, $a= 0$, $b = \pi/8$, $\Delta^2 = 0.1$. Right: number of coincidences.

In summary: the time evolution of the efficiency in each station follows the time scale of the other. If the setup is only slightly asymmetric, the spontaneous synchronization effect holds (otherwise, it doesn't). If there are two widely different timescales, the system relaxes first with the "fast" one, and then evolves according to the "slow" one. The general problem is far more complex that the symmetrical case. The possibilities of new types of observable predictions (as the mentioned variation of efficiency with the angle settings) are vast and promising, and remain to be fully explored.

## 5. Results obtained in an incomplete experiment.

In the year 2012, our group tested and closed the *time coincidence loophole* [24,25]. That experiment can also be regarded as an incomplete version of the one proposed in Section 3 here. It involved a pulsed source of entangled photons and a time-stamped record of both photon detection and pulse emission events, as required, but it didn't fit the time hierarchy described in *(iii)*. In particular, $\tau_{res} \gg \tau$, and the pump pulse was not square-shaped. Most important, the values of pump intensity, $\tau_{pulse}$, $L$ and $R_p$ were not varied. Anyway, it is sensible to have a look at the obtained results. As it is shown next, they are consistent with AB. It is important to stress here that we do not claim them to provide any sort of confirmation of AB, for some artifact might have been their cause (unfortunately, the setup was dismantled). We think they just encourage the realization of the appropriate experiment, and provide a useful example of how real data look like.

In Figure 7, measured $\eta(t)$ and $S_{CHSH}(t)$ during the pump pulse are shown. They are obtained from the set of 36 files lasting nearly 56 s of real time each, recorded for equally spaced values of $\{a,b\}$, that were used in [23] to measure the time variation of the Concurrence. As it was already observed in [23,25], quantum correlation is constant during the whole pulse. This result agrees with predictions of both QM and AB. Efficiency, instead, increases almost linearly, starting with ≈3% and reaching almost 16% some 100 ns (or ≈400 $\tau$) later. This variation is not predicted by QM and is therefore caused by an artifact or by an effect of the kind described by the AB model. The curve of $<\eta>(t)$ in Fig.7 sums up coincidences recorded for all $\{a,b\}$ settings in order to improve the statistics. The same variation (i.e., nearly linear, and with the same slope) is observed in each subset of data corresponding to each setting $\{a,b\}$.

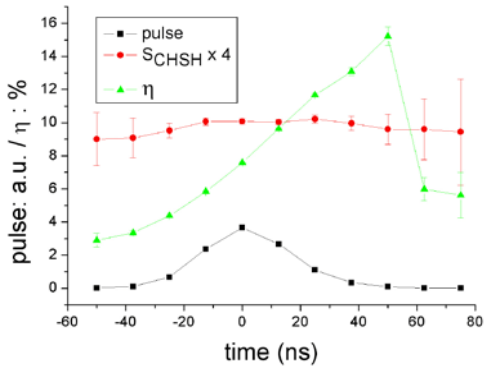

Figure 7: observed $<\eta>(t)$ (in %) and $S_{CHSH}(t)$ (×4, for clarity of figure) in the 2012 experiment [25], $\tau_{res}$=12,5 ns, $\tau$=0,27 ns, $R_p$ =60 KHz, station B. $S_{CHSH}$ at pulse peak is 2,52 ± 0,04. Average rate of coincidences with analyzers removed is 300 s$^{-1}$. Single counts (average 2000 s$^{-1}$) are also plotted (in arbitrary units), to indicate pulse duration (35 ns FWHM), shape and position. The best fit to the (assumed linear) slope of $<\eta>(t)$ is 1,2×10$^{-3}$ ns$^{-1}$ (3,2×10$^{-4}$ $\tau^{-1}$), $<\eta>$(t=-50ns)= 0,029.

Observable efficiency is affected by well-known technical imperfections, being the most important ones: quantum yield of photodiodes, transmission of spectral filters and precision of alignment. In our case, these imperfections reduce the ideal value in a factor about 3. Taking into account this factor to correct the measured values and using eqs.16-17, Fig.7 is consistent with AB if $\Delta$≈0,06 and $\Gamma$≈0,04 ns$^{-1}$ (≈0,01 $\tau^{-1}$). Note that $\tau_{res} \gg \tau$, so that oscillations, if existed, would have not been observed. This limitation leaves space to additional sets of values of $\{\Delta,\Gamma\}$ to be consistent with Fig.7. Yet, these values can be found only through numerical simulation. If $\Delta$ is assumed to be the same, the observed slope is fitted with $\Gamma$≈ 1,57 $\tau^{-1}$.

Before -50 ns and after 50 ns the pump pulse intensity almost vanishes, the number of coincidences is low, and hence error bars are large. Estimated $<\eta>$ at -62,5 ns (not plotted) follows the same trend than later values. Instead, $<\eta>$ at +62,5 and +75 ns drops abruptly, suggesting a decay time $\tau_d$ ≈ 25ns ≈ 100 $\tau$. On the other hand, $S_{CHSH}$ (also the Concurrence) remains constant even in the large-error region, in agreement with both QM and AB.

Fig.7 is drawn from the best set of data recorded in the 2012 experiment. Similar results were obtained in repetitions of the experiment performed in 2018 using the same setup. They also show a linear variation of $<\eta>$(t), but with different (yet not too different) slope. There is no reason for the change of slope according to the AB model, what suggests the observed linear variation to be caused by some sort of instrumental imperfection. However, this hypothetical imperfection leaves $S_{CHSH}$ and Concurrence unchanged. Its nature is hence not obvious. At this point, we believe that investing efforts in performing the experiment proposed in the Section 3 is more profitable than trying to figure out the cause of the imperfection in the performed, incomplete experiments.

**5. Summary.**

AB is a LR model able to reproduce all QM predictions after a time >$L/c$ (after turning on the source of entangled states) has elapsed. It has not been refuted by experiments performed until now, and can be tested in a simple setup. The magnitude to measure is time variation of efficiency, and how this variation is affected by the distance between stations, in a "stroboscopic" version of the experiment in Fig.1. Data recorded in a similar (but incomplete) experiment performed in 2012 and repeated in 2018 are consistent with AB. Yet, these data are insufficient to validate it.

The observations described in *I)* and *II)* in Section 3 define a test of QM vs LR of a new type, independent of the violation of an average correlation limit. In the usual tests (i.e., the violation of a Bell's inequality) threshold values of detectors' efficiencies and angle settings' unpredictability, as well as event-ready signals, must be ensured to close all the known loopholes. In the test proposed here, instead, a definite answer is obtained even if those difficult technical requisites are not fulfilled. In the usual tests, a low correlation, say, $S_{CHSH} < 2$, cannot be interpreted as a refutation of QM, for it may be the consequence of a poor realization. In the test proposed here, instead, the observation of *I)* and *II)* would refute QM (as we know it) even in an imperfect setup. The case of the strongly asymmetrical setup offers the possibility of additional observable predictions, as the dependence of the efficiency with the angle settings.

The usual tests face the problem of the never-ending proliferation of loopholes (and hence technical conditions increasingly difficult to achieve), never reaching fully conclusive results. The experiment proposed here provides a definite answer for both possible outcomes: if $L$-dependent $<\eta>(t)$ dynamics are observed, the usual interpretation of QM is disproved, and the door to potentially fruitful nonlinear generalizations of QM are open. If such $L$-dependent dynamics are not observed instead, the hypothesis underlying AB is disproved.

Finally, the experiment is technically simple: the analyzers can be kept fixed and efficiency and correlation are not needed to reach a threshold. We believe it difficult to find an experiment at hand with more important potential consequences.

**Acknowledgments.**

Many thanks to Peter Morgan (Yale University), Siddarth Joshi and Rupert Ursin (IQOQI) for their interest in this work, their perceiving questions, observations and advices. This work received support from the grants N62909-18-1-2021 Office of Naval Research Global (USA), and PIP011-077 and PIP2017-027C from CONICET (Argentina).